\documentclass[preprint,12pt]{elsarticle}
\pdfoutput=1 
\usepackage{epsfig}
\usepackage{here}
\usepackage{graphicx}
\usepackage{amssymb}

\usepackage{lineno}
\usepackage{comment}
\usepackage{color}
\usepackage{amsmath}

\begin{document}

\begin{frontmatter}



\title{The development of radioactivity measurement system in gases}

\author[Nihon]{H.~Ogawa}

\address[Nihon]{Department of Physics, College of Science and Technology, Nihon University, Kanda, Chiyoda-ku, Tokyo 101-8308, Japan}

\begin{abstract}
A device for measuring radioactive impurities in gas is developed, consisting of an electrostatic collection type radon detector and an extremely low radioactivity proportional counter. This device measures the radioactive impurities released by materials into the gas and a filtering test for radon removal from gas. The performance of the newly developed proportional counter was evaluated by measuring radon rate. The calibration factor of $^{222}$Rn as 27.4$\pm$0.2~count/day/mBq and the background rate of $^{222}$Rn as 0.84$\pm$0.13~mBq were estimated. 
\end{abstract}




\end{frontmatter}


\section{Introduction}
To improve the sensitivity of direct dark matter search experiments, it is critical to accurately evaluate and reduce the radioactive impurities contained in the dark matter search detector. Noble gases such as xenon and argon are common targets in the search for these experiments. Radioactive impurities in these gases serve as the background for dark matter exploration experiments. Particularly, radioactive radon gases quantitatively emanate from detector materials; thus, their measurement and removal has an important research subject. 

The content of radioactive radon in the gas in the future large dark matter search detectors will be 0.1~$\mu$Bq/kg (weight of dark matter search target) at a sensitivity of dark matter scattering cross-section O (10$^{-49}$) cm$^{2}$~\cite{DARWIN}. Until now, such levels of radioactivity had to be evaluated from the background of a dark matter search detector. In this case, it may not be possible to distinguish between signals from dark matter and radioactive backgrounds. Therefore, it is neccesary to estimate the background independently of the dark matter search detector.

In this study, a device for measuring radioactive impurities in gas, which is the target of dark matter search experiments, using an electrostatic collection type radon detector and an extremely low radioactivity proportional counter is reported. This device measures the radioactive impurities released from materials into the gas and a filtering test of radon removal from gas. An electrostatic collection type radon detector has been widely used to measure radon in the air and water in the Super-Kamiokande experiment~\cite{Hosokawa, Pronost}. 
In addition, a study was carried out to measure the radioactivity in gases using ultra-trace radioactivity proportional counters.

The contents of the radioactivity measurement system are explained in Sec.~\ref{sec:overview}, and the performance of the proportional counter is indicated in Sec.~\ref{sec:prop}. Finally, the conclusion is presented in Sec.~\ref{sec:concl}.

\section{Device overview}
\label{sec:overview}
\begin{figure}[htbp]
  \begin{center}
    \includegraphics[keepaspectratio=true,height=80mm]{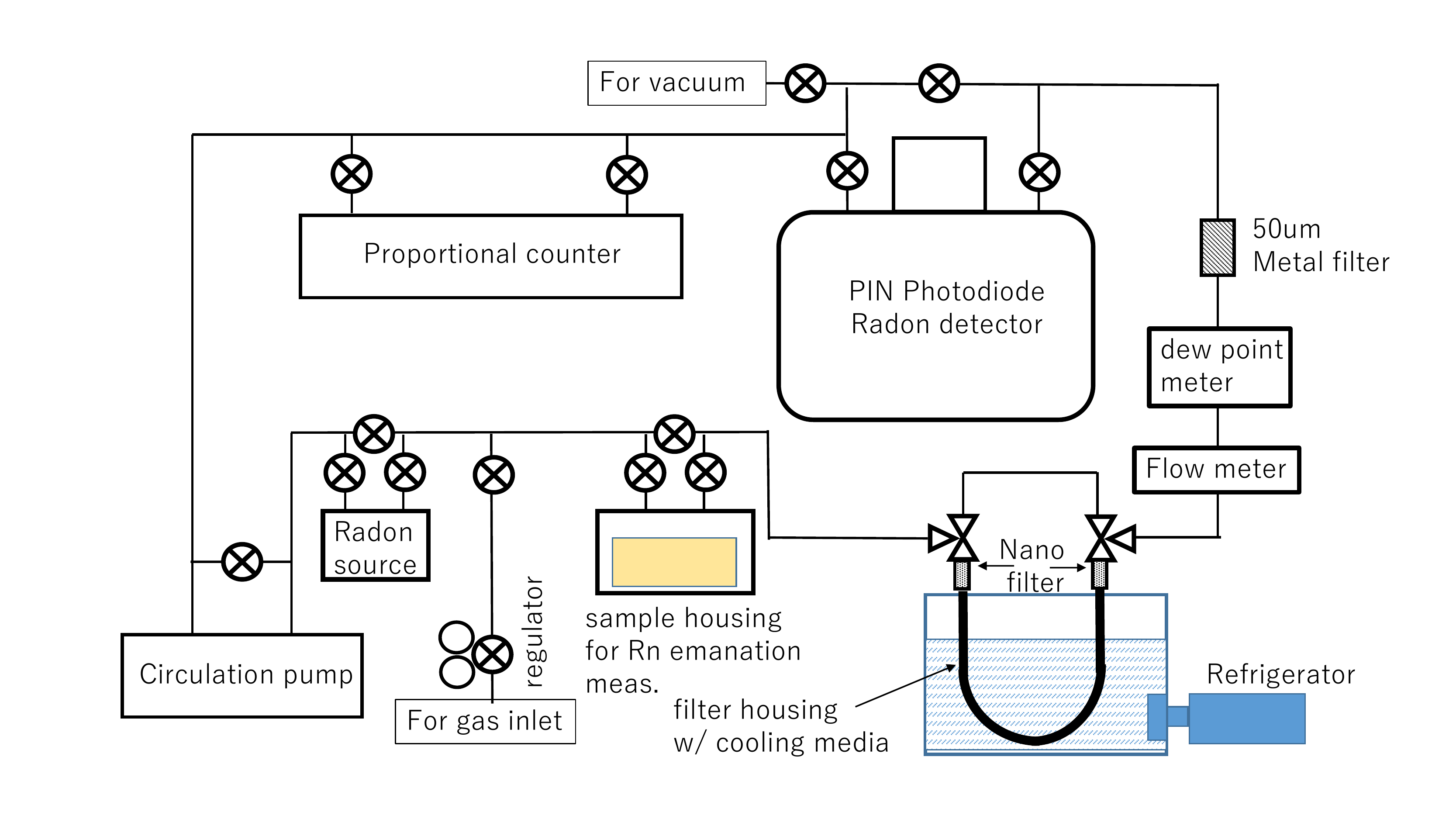}
  \end{center}
  \caption{The setup of $^{222}$Rn detector and proportional counter}
  \label{fig:setup1}
\end{figure}

Figure~\ref{fig:setup1} shows the radioactivity measuring device used in this study. This device was developed at the Nihon University College of Science and Technology Funabashi Campus. The device consists of an electrostatic collection type radon detector (PIN photodiode radon detector) and a proportional counter, which are connected via a gas pipe, and the inside is filled with gas. The device can circulate gas with a circulation pump and measure the radioactivity in the gas in real-time. The device also includes a sample container and a filter housing. In the former, a material can be inserted to measure the discharge of radioactivity (mainly radon) from material. The latter can contain an adsorbent and be cooled. Then, the adsorption test of radioactivity in the gas can be performed. Here, the details of each device are explained.
\subsection{Electrostatic collection type radon detector}
The electrostatic collection type radon detector has been widely used as a device for measuring radon in the air in the Super-Kamiokande experiment~\cite{Hosokawa, Pronost}. The device container is made of stainless steel with a capacity of 80-L (litter), and the inner surface is electrolytically polished. The $^{214}$Po, the daughter nucleus of radon, is collected using a high electric field, and alpha rays emitted from the decay of $^{214}$Po are detected with high resolution using a photodiode. The detection sensitivity of radon is about 0.1~mBq which is estimated by the background measurement of the radon detector.

\subsection{Proportional counter}
\begin{figure}[htbp]
  \begin{center}
    \includegraphics[keepaspectratio=true,height=80mm]{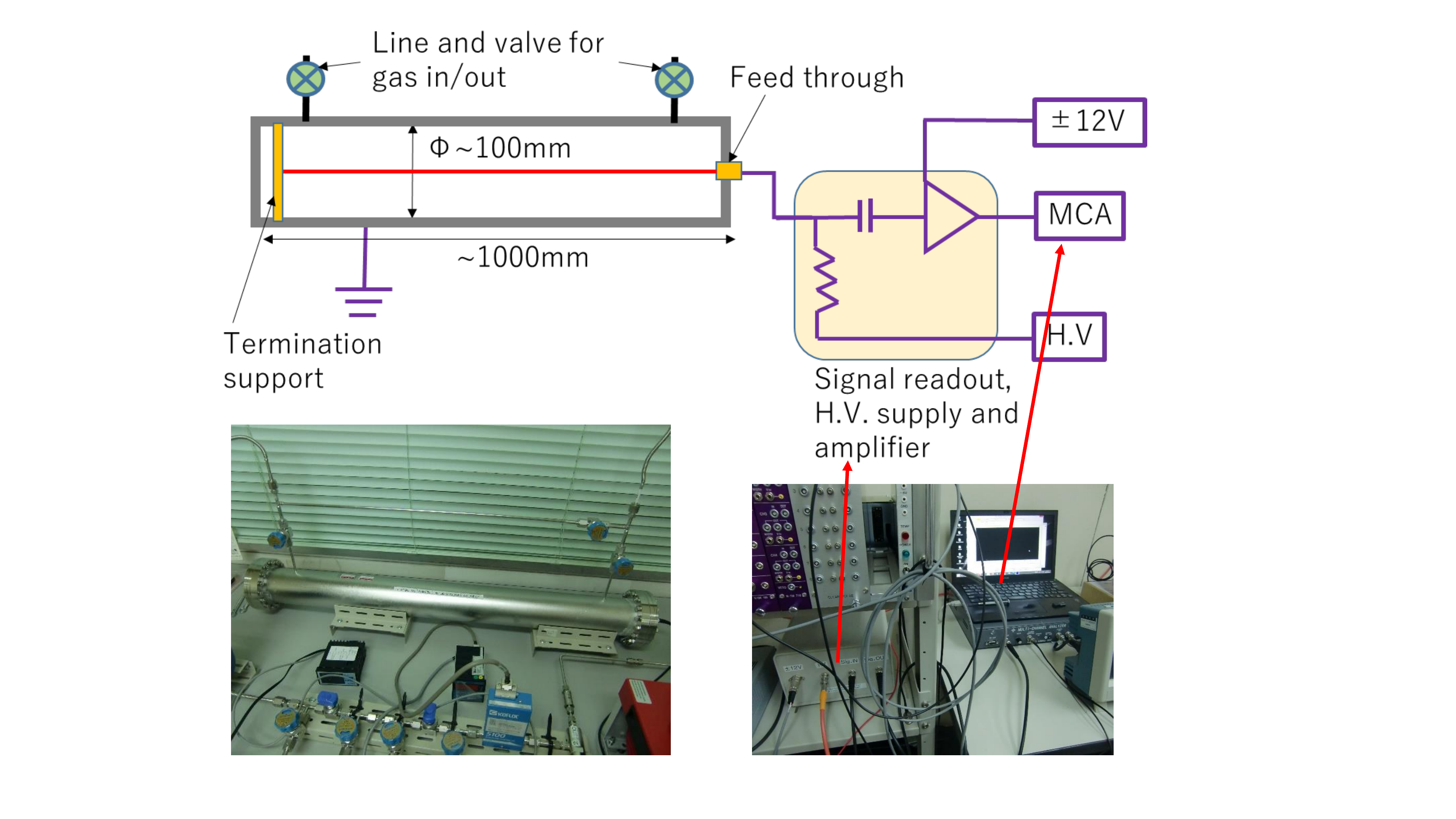}
  \end{center}
  \caption{The setup of proportional counter}
  \label{fig:setup2}
\end{figure}

Figure~\ref{fig:setup2} shows a schematic of the proportional counter setup and an overall image. The proportional counter body is made of stainless steel and has a large capacity of a 100-mm diameter and 1000-mm length. The total volume of gas region is about 8-L. The inner surface of the main body was also electropolished. The core wire is a 30-$\mu$m gold-plated tungsten one. The feed through that holds the core wire and the insulator of the terminate support are made of Kyocera A500 low-radioactive alumina. Circuits for signal decoding, high voltage (HV) supply, and amplifiers have been designed for this proportional counter. The signal from the proportional counter is read using pulse height analyzer (PHA) to derive the energy distribution. This PHA has a repeater function that can be used to investigate time dependence. 

The proportional counter technology used in this application is existing one. A previous study at the Pacific Northwest National Laboratory was conducted on the measurement of radioactivity in gas using an ultra-low radioactivity proportional counter~\cite{EKMace}, which is mainly used for determining the content of isotope $^{39}$Ar in argon gas~\cite{Ar39}.  In this study, it is possible to estimate the amount of radon in gas more directly by directly observing alpha rays from $^{222}$Rn with a proportional counter. By combining these two radioactivity measuring instruments, we can measure low radioactivity radon independently with high sensitivity while having a relatively compact size. 

\subsection{Filter system and housing for material measurements}
This device can measure radon emanation from materials and perform radon adsorption tests on gas. Material can be placed inside a 10-L stainless steel container prepared for the former. The latter also has a filter and cooling system. Metal particle filters are placed before and after the adsorption filter to prevent the filter members reaching the detector. A filter housing and refrigerant are inserted in a Dewar bottle, and the filter can be cooled to about -60 $^\circ$C by the refrigerator. A thermometer is installed in the filter part to control the temperature of the refrigerator.

\subsection{Other devices in the gas circulation line}
The gas was circulated using a diaphragm pump. The gas contact part of this pump is made of metal and is vacuum resistant to prevent impurities from entering from the outside. A gas flow meter and a dew point meter are installed as monitors in the gas line. The gas line is equipped with a vacuum drawing port, a gas introduction port, and a gas recovery port. A radon source with radium ceramic balls was also provided for detector calibration and radon removal test. A noise cut transformer is installed in the power supply of the measuring device, and the measuring device, measuring device and other devices are connected to the ground. 

\section{Performance of proportional counter}
\label{sec:prop}
\subsection{Detector adjustment}
\begin{figure}[htbp]
  \begin{center}
    \includegraphics[keepaspectratio=true,height=50mm]{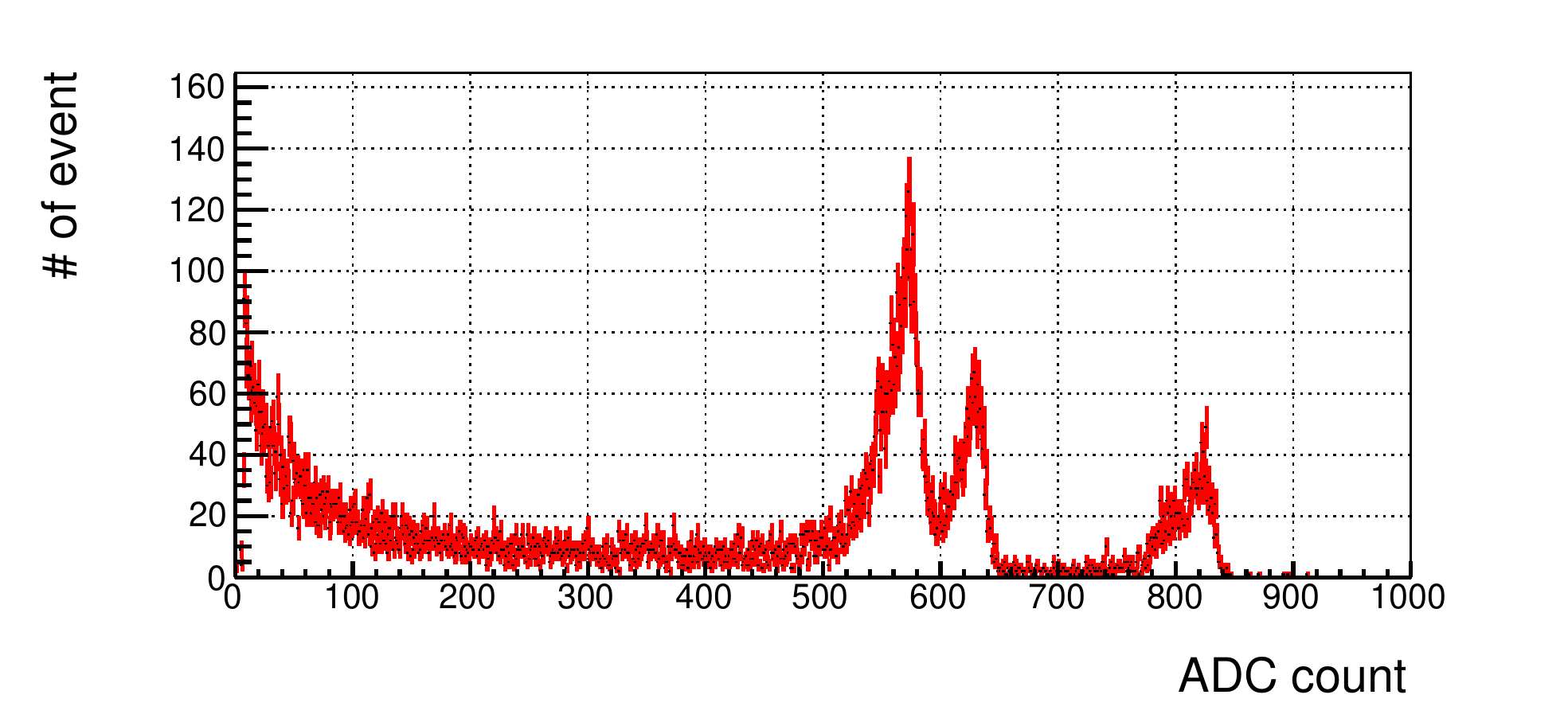}
  \end{center}
\caption{(Left) The PHA channel distribution observing alpha rays from radon decay. The applied voltage value is 1000~V.} 
\label{fig:spec1}
\end{figure}

In investigating the performance of the proportional counter 1~atm of argon gas was introduced after evacuating the inside of the system. Argon in the system was then pumped through a vessel containing radium ceramic balls that released radon. An HV of the proportional counter was applied, and the signal from the alpha ray generated by the decay of the radon series was observed using PHA. The PHA channel distribution observing alpha rays from radon decay when the applied voltage is 1000~V is shown on the left of Fig.~\ref{fig:spec1}. Alpha rays from the $^{222}$Rn, $^{218}$Po, and $^{214}$Po collapses are observed.  

\begin{figure}[htbp]
  \begin{center}
    \includegraphics[keepaspectratio=true,height=70mm]{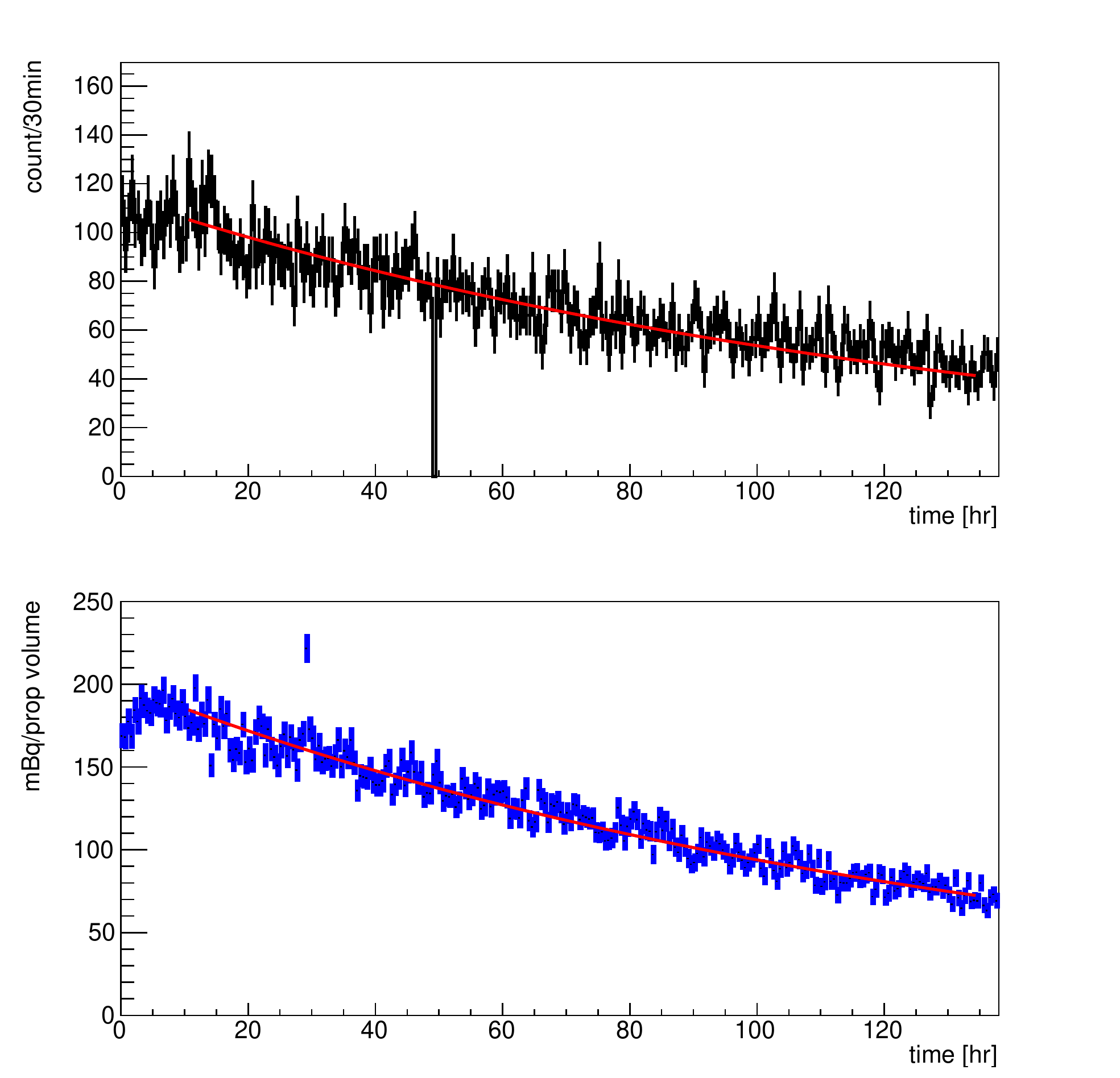}
  \end{center}
  \caption{Time variation of radon concentration in argon for the proportional counter (top) and radon detector (bottom) with the fitting curve.}
  \label{fig:radoneff}
\end{figure}

The efficiency of radon detection was calibrated by circulating radon injected argon gas in series with an electrostatic radon detector and comparing both count rate. The time variation of radon concentration in argon for the proportional counter and radon detector is shown in Fig~\ref{fig:radoneff}. The count rate of $^{222}$Rn is estimated using the ROI 500--590 of ADC count. The radon rate in the radon detector was estimated using the correction factor of the type of radon detector shown in~\cite{Hosokawa}. The decay function of the radon with a lifetime $\tau$ (half-life of $^{222}$Rn 3.82~days/$ln2 =$ 5.82~days) was used to fit the time variation of the both rates as. 
\begin{equation}
R = A\cdot e^{-t/\tau}
\end{equation}
$t$ is the time since measurement and $A$ is the event rate in the proportional counter or radon rate in the radon detector. The $A$ is estimaated as 114.0$\pm$ 0.9 count/30~min for the proportional counter and 199.9$\pm$0.7 mBq for the radon detector which the volume is scaled to one of the proprtional counter.  
Then, the count rate of the proportional counter was scaled to the assayed activity of the radon detector. 1~mBq for the proportional counter is corresponding to 0.570$\pm$0.005 count/30~min. The calibration factor for the density of radon in the proportional counter is estimated as 27.4 $\pm$ 0.2 count/day/mBq. 
In addition, this measurement confirmed that the gain of the signal was stable over 5 days confirmed by the stable peak of the ADC count for the signal from alpha-rays.

\subsection{Detector background}
\begin{figure}[htbp]
  \begin{center}
    \includegraphics[keepaspectratio=true,height=70mm]{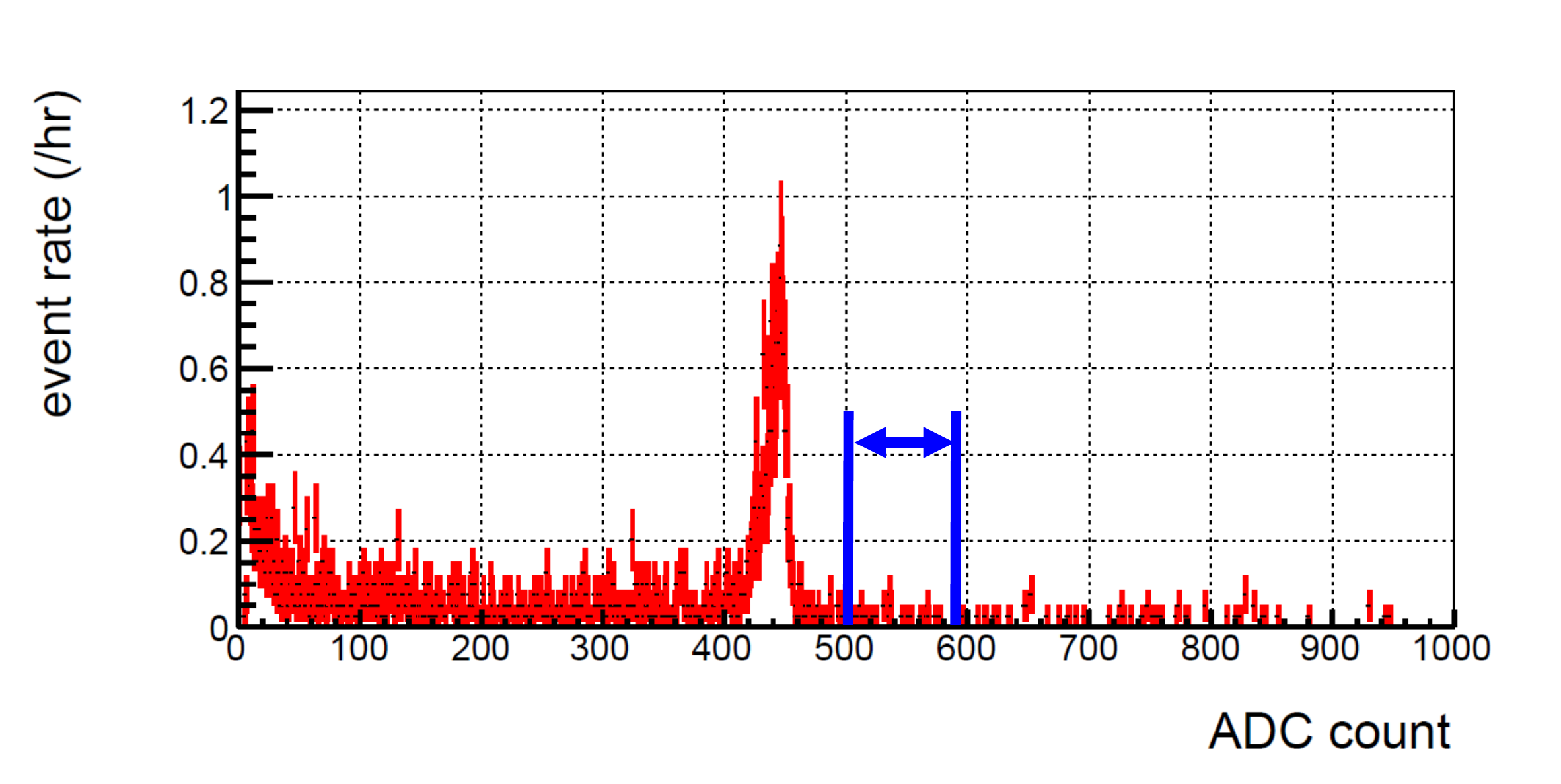}
  \end{center}
  \caption{Background energy spectrum in the proportional counter. Blue region indicates the ROI for the $^{222}$Rn background estimation.}
  \label{fig:BG}
\end{figure}

Then, the background of the proportional counter is measured for 40~hours. The background energy spectrum of the proportional counter is shown in Fig.~\ref{fig:BG}. The peak around 440 analog-to-digital converter (ADC) count is due to alpha from $^{210}$Po decay, which applied the surface of the proportional counter. The radon concentration in this background data is estimated using the same range of the ROI 500-590 of the ADC count as 23.1$\pm$3.7~count/day, corresponding to 0.8 4$\pm$0.13~mBq.  

\section{Conclusions}
\label{sec:concl}
A device for measuring radioactive impurities in gas is newly developed, consisting of an electrostatic collection type radon detector and an extremely low radioactivity proportional counter. The performance of a newly developed proportoinal counter is evaluated. Long-term stable data acquisition over multiple days was confirmed. The calibration factor of $^{222}$Rn is estimated as 27.4$\pm$0.2~count/day/mBq. The background rate is about 0.84$\pm$0.13~mBq.

\section*{Acknowledgments}
This work was supported by the Japanese Ministry of Education, Grant-in-Aid for Scientific Research on Innovative Areas Region No. 2603, JSPS KAKENHI Grant No. 19K03893, Shingakujyutsu 19K05806, Research grants Foundation for Precision Measurement Technology (2019), Research grants Foundation for CST Nihon University (2020), Research grants Foundation for the Yamada Science Foundation (2021-2022) and Academic Award Scholarship for CST Nihon University (2022).




 
\end{document}